\documentclass[twocolumn]{aastex63}

\shorttitle{Clumpy accretion of UX~Ori type stars}
\shortauthors{Grinin \& Demidova}

\begin{document}

\title{Clumpy accretion as a possible reason of prolonged eclipses of UX~Ori type stars}
\correspondingauthor{Tatiana V. Demidova}
\email{proxima1@list.ru}

\author[0000-0001-8923-9541]{Vladimir P. Grinin}
\affiliation{Pulkovo Observatory of the Russian Academy of Sciences, Pulkovskoje Avenue 65, St. Petersburg 196140, Russia}
\affiliation{V.V. Sobolev Astronomical Institute, St. Petersburg University, Petrodvorets, Universitetskiy p. 28, St. Petersburg, 198504, Russia}

\author[0000-0001-7035-7062]{Tatiana V. Demidova}
\affiliation{Crimean Astrophysical Observatory,p. Nauchny, Bakhchisaray, Crimea, 298409, Russia}

\begin{abstract}
The paper proposes a model of deep and prolonged eclipses of young stars of the UX~Ori type. Some of these events continue for decades and existing models cannot explain them. Our paper shows that such eclipses can be caused by the falling of gas and dust clump from the remnants of the protostellar cloud onto the protoplanetary disk. The disturbance in the disk caused by the fall of the clump leads to a surge in the accretion activity of the star and, as a consequence, to an increase in the disk wind. If the circumstellar disk is tilted at a slight angle to the line of sight, then dust raised by the wind from the surface of the disk can cause a strong decrease in the star's brightness, which can last for decades.
\end{abstract}


\section{Introduction}\label{sec:intro}
UX~Ori type stars are among the photometrically most active young objects. The reason for their activity is the slight inclination of the circumstellar disks relative to the direction towards an observer~\citep{1991Ap&SS.186..283G, 1997ApJ...491..885N}. As a result, the brightness of these stars changes following the change in circumstellar extinction along the line of sight. The amplitude of brightness reduction in the optical region of the spectrum reaches 2-3$^m$. The decrease in the star's brightness below this level is prevented by the scattered radiation of its circumstellar disk, which dominates at deep minima~\citep{1988SvAL...14...27G}. The duration of minima usually ranges from several days to two - three weeks. Such events can be caused by fluctuations in the density of the gas-dust atmosphere of the disk in the dust evaporation zone~\citep{2001A&A...371..186N,2001ApJ...560..957D} or the inhomogeneous structure of the disk wind~\citep{2007ApJ...658..462V,2008AstL...34..231T}, or clouds of charged particles rising above the disk plane along the magnetic field lines~\citep{2014ApJ...780...42T}. However, in some cases very long eclipses are observed, lasting ten or more years~\citep[see, for example,][]{2015A&A...582A.113S,2023Ap.....66..235G}. Such events still have no acceptable explanation. Our paper proposes a semi-empirical model of such events. It is assumed that they may be the result of massive gas and dust clumps falling onto the disk from the remnants of a protostellar cloud or from circumstellar space.

The theory predicts the formation of such clumps in the vicinity of protostars at the early stages of their evolution~\citep{2018MNRAS.475.2642K}. The accretion of such clumps can occur on the inner parts of the disk ($\sim 1$~a.e). In addition, clumps weighing several tens of Jupiter masses can form due to the development of gravitational instability at the periphery of massive protoplanetary disks~\citep{2005ApJ...633L.137V}. Some of these clumps lose angular momentum as they interact with the surrounding gas and migrate toward the center of the disk, causing bursts of accretion. In addition, clumps can leave the parent disk during the process of rapid relaxation of the system~\citep{2012ApJ...750...30B,2017A&A...608A.107V}. The ejection of clumps will help stabilize the system by analogy with the processes occurring in young star clusters~\citep[for example,][]{1998MNRAS.294...47G}.

Such gas clumps can be observed as compact dark spots against the background of bright reflection nebulae and HII regions~\citep{2007AJ....133.1795G,2018Ap&SS.363...28G}. When a compact clump approaches a protoplanetary disk, it will be stretched into a stream of finite size. Asymmetric stream-like gas structures have been discovered in high-contrast images of a number of protoplanetary disks obtained with ALMA~\citep{2023ASPC..534..233P}. The most striking examples of such structures were observed in the young objects SU Aur~\citep{2021ApJ...908L..25G} and DG~Tau~\citep{2022A&A...658A.104G}. This suggests that the accretion of gas clumps onto a protoplanetary disk from the surrounding space is not an exceptionally rare event. 

In our previous papers (\citet{2022ApJ...930..111D,2023ApJ...953...38D} using the SPH method, three-dimensional gas-dynamic models were calculated, on the basis of which the observational manifestations of the consequences of a collision of a gas clump with a protoplanetary disk were studied. It has been shown that when observing the disk from polar directions, the fall of the clump leads to the formation of large-scale structures that can be observed in submillimeter images. In this case, the fall of a massive clump near the star can trigger an FU Ori-type flare, as well as tilt the inner part of the disk relative to the periphery. In this paper, we study the case when the protoplanetary disk is observed at a small angle to the line of sight (the case of UX~Ori type stars).

\section{Model}\label{sec:mod}
The paper considers a model of a system consisting of a star with a mass of $M_\ast=1.4 M_\odot$, a radius of $R_\ast=2.3 R_\odot$ and a temperature of $T_\ast=7500K$, surrounded by a gas disk. At the initial moment of time, the matter in the disk is distributed within the limits $r_{in}=0.1$~au,$r_{out}=50$~au according to the standard density distribution law described in ~\cite{1994A&A...286..149D}: 
\begin{equation}
\rho(r,z,0)=\frac{\Sigma_0}{\sqrt{2\pi}H(r)}\frac{r_{in}}{r}e^{-\frac{z^2}{2H^2(r)}},
\label{Eq:rho}
\end{equation}
where $r$ is the distance from the center of the star in the disk plane (cylindrical radius), and $z$ is the distance from the disk plane, $\Sigma_0$ is the surface density at a distance $r_{in}$ (determined by the total mass of the disk $M_{disk} = 0.01M_\odot$). The half-thickness of the disk is determined by the formula:
\begin{equation}
H(r)=\sqrt{\frac{\kappa T_{mid}(r) r^3}{GM_{\ast} \mu m_H}},
\end{equation} 
where $T_{mid}$ is the temperature of the matter in the plane of the disk, $G$ is the gravitational constant, $m_H$ is the mass of hydrogen, and $\mu=2.35$ is the average molecular mass.

The disk is assumed to be vertically isothermal. The dependence of temperature on the distance to the star is described by a relation similar to that adopted in ~\cite{1997ApJ...490..368C,2004A&A...421.1075D}:
\begin{equation}
T_{mid}(r)=\sqrt[4]{\frac{\Gamma}{4}}\sqrt{\frac{R_{\ast}}{r}}T_{\ast},
\label{temp}
\end{equation}
where the parameter is $\Gamma=0.05$, and does not change over time.

\begin{figure*}[ht!]
\centering \includegraphics[width=1.0\textwidth]{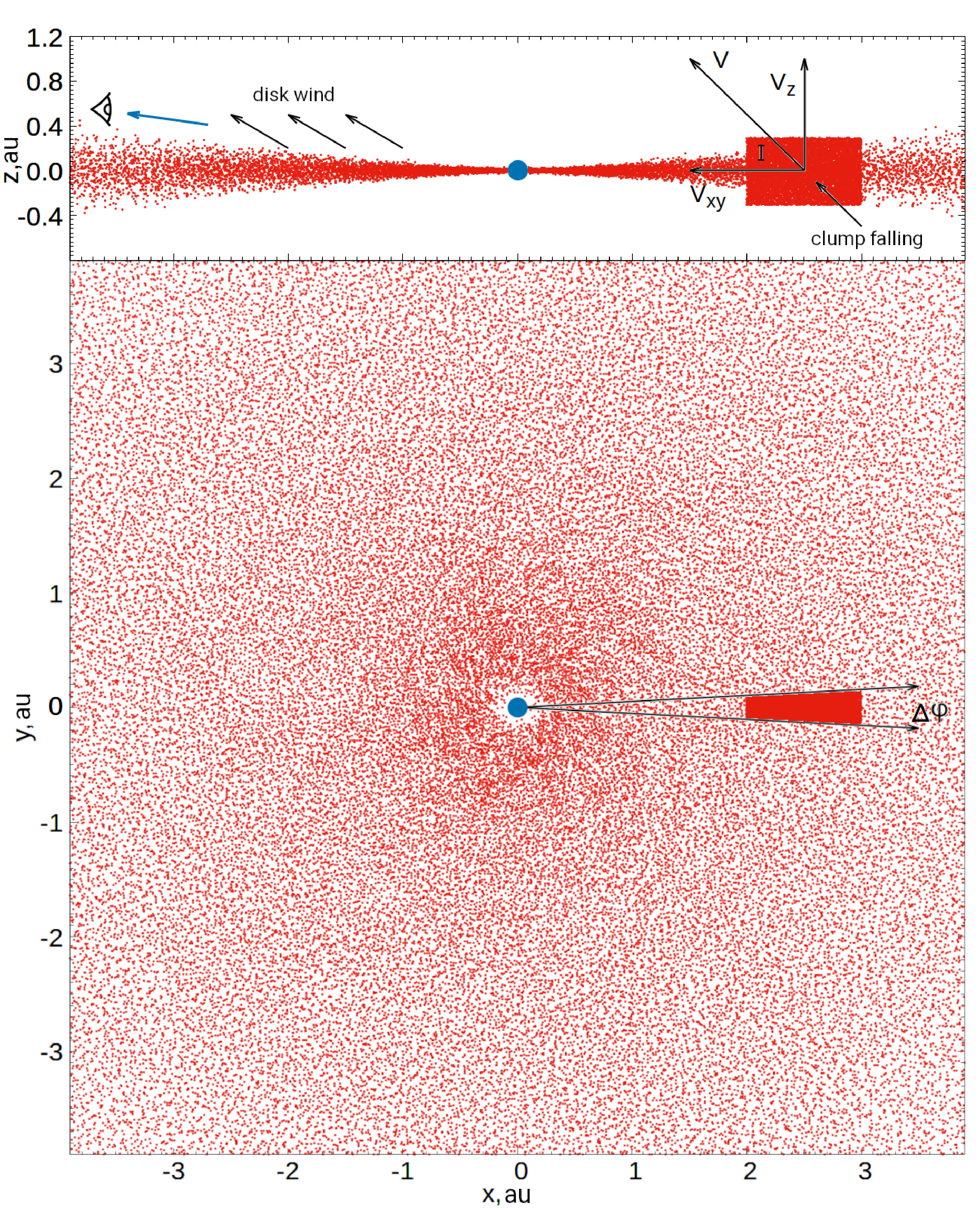}
\caption{\normalsize The section of the disk in the $xz$ plane after adding one portion of the clump substance (top). And the projection of the position of gas particles onto the $xy$ plane (bottom) for the model $m=0.3M_{Jup}$, $r_0=2$~au, $r_1=3$~au, $L= 0.5$, $I=45^\circ$, $\Delta\phi=6^\circ$.}\label{fig0}
\end{figure*}

The dynamics of the disk were calculated over 600 years, and then a density perturbation was added to it in portions (which simulates the fall of a gas steam) within the radii $r_0$, $r_1$ and the azimuthal angle $\Delta\phi$ (Fig.~\ref{fig0}). A total of five portions of gas were added with a time interval of $0.2$~year (calculations showed that during this time a point at a distance of $r=3$~AU, located near the apoaster of the orbit, passes $\sim 6^\circ$). It was believed that during the collision with the disk, the matter lost some of its energy to heat the disk. Therefore, it was assumed that the perturbation velocity is less than the Keplerian one and is a fraction of $L$ from it at a given distance (in addition, when the clump falls retrogradely, its speed will decrease due to interaction with the disk matter). The vertical component of the velocity, corresponding to the inclination $I$, was also preserved, since it was assumed that the clump’s orbit intersects the plane of the disk at a certain angle. The perturbation mass $m$ also varied. A detailed description of the model is given in~\cite{2023ApJ...953...38D}. After adding the clump, calculations continued for another 500 years.

\section{Method}\label{sec:method}
Gas-dynamic calculations were carried out using the SPH method with variable smoothing length, which is the basis of the cosmological code GADGET-2\footnote{\url{https://wwwmpa.mpa-garching.mpg.de/gadget/}}~\citep{2001NewA....6...79S,2005MNRAS.364.1105S}, modified by us for protoplanetary disks~\citep{2016Ap.....59..449D}. $10^6$ particles with gas properties took part in the calculations. The viscosity and self-gravity of the disk were taken into account.

In the calculations, the transfer of angular momentum was carried out by introducing numerical viscosity, for which the following parameters were set: $\alpha_{SPH}=1$, $\beta_{SPH}=0$. Turbulent viscosity in this case is defined as $\nu=0.1\,c_s\,h\,\alpha_{SPH}$~\citep{2012JCoPh.231..759P}, where $c_s$ is the speed of sound, defined within a sphere limited by the smoothing length $h$ in the SPH method. Thus, the Shakura-Sunyaev viscosity parameter $\alpha_{SS}$~\citep{1973A&A....24..337S} is not constant and is related to $\alpha_{SPH}$ by the relation: $\alpha_{SS} =0.1\alpha_{SPH}\frac{h}{H(r)}$. In an unperturbed disk, the value of $\alpha_{SS}$ varies from $0.36$ to $0.01$.
\newpage
\section{Results}\label{sec:res}
\subsection{Eclipse of a star by the disk wind}\label{sec:res1}

\begin{figure*}[ht!]
\centering \includegraphics[width=0.8\textwidth]{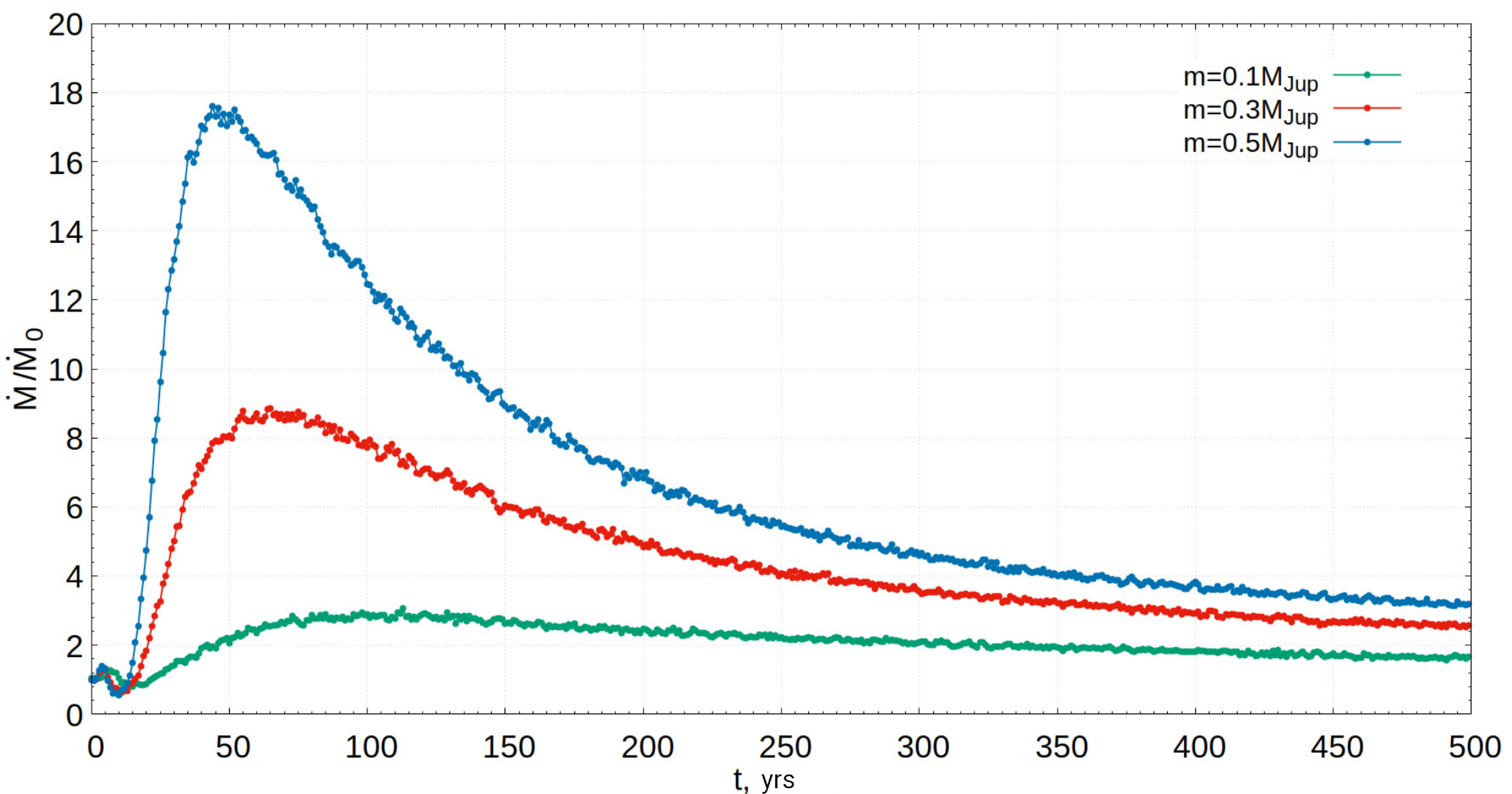}
\caption{\normalsize Accretion rate as a function of time for three values of the initial clump mass: $0.1M_{Jup}$ (green line), $0.3M_{Jup}$ (red line) and $0.5M_{Jup}$ (blue line). The value is given relative to the unperturbed value $\dot{M}_0$. Model parameters are $r_0=2$~au, $r_1=3$~au, $L=0.5$, $I=45^\circ$, $\Delta\phi=6^\circ$ .}\label{fig1}
\end{figure*}

The figure~\ref{fig1} shows a surge in the accretion rate caused by the fall of a gas clump from the circumstellar environment of the star onto the disk at the moment $t = 0$. The place where it falls onto the disk is located at a distance of $2-3$~au from the star. An accretion burst begins at the moment when a wave of disturbances in the disk caused by the fall of a clump reaches the immediate vicinity of the star. At the flare maximum, the accretion rate exceeds that in the unperturbed state of the disk ($\dot{M}_0=1.3\cdot 10^{-7} M_\odot$/yr) by approximately $3$ times in the case of $m=0.1M_{Jup }$ ($M_{Jup}$ is mass of Jupiter), $9$ times for $m=0.3M_{Jup}$ and $18$ times for $m=0.5M_{Jup}$. In addition, with an increase in the initial mass of the clump, the time between the beginning of the increase in the accretion rate and the achievement of its maximum value decreases.

Calculations have shown that turbulent viscosity in the region $r<3$ a.u. decreases to $\frac{2}{3}\nu_0$ ($\nu_0$ --- unperturbed value) after 10 years, after adding a disturbance to the gaseous medium of the disk, and then slowly increases and after 100 years it is $0.7\nu_0$ . Thus, the increase in the accretion rate in the model is associated primarily with a significant increase in the amount of matter with sub-Keplerian velocities in the region under consideration, and not with a change in viscosity.

It should be noted that the flare of accretion activity of the star caused by the fall of the clump onto the disk should also be accompanied by an increase in infrared (IR) radiation from the disturbed region of the disk. The shape and amplitude of the IR response depend on the model parameters and the angle of inclination of the disk to the line of sight. An additional source of IR radiation can also be a dusty disk wind (see below).

It is assumed that the surge in accretion activity is accompanied by an increase in the magnetic-centrifugal disk wind. In the case of FU~Ori type stars, an increase in the disk wind was observed directly from the spectra of the star~\citep{1996ARA&A..34..207H}. It is known from theory~\citep[see, for example,][]{2000prpl.conf..759K} that the rate of mass loss in a magnetic-centrifugal wind $\dot{M}_{wind}$ and the accretion rate $\dot {M}$ are closely related: $\dot{M}_{wind} \approx 0.1 \dot{M}$. During the acceleration process, the wind material involves small dust particles and lifts them above the surface of the disk~\citep{1993ApJ...408..115,2017ApJ...845...44T}. As a result, the disk wind can become a source of circumstellar extinction. The optical thickness of the wind in the direction toward the observer $\tau$ depends on the type of wind model and the inclination of the disk and is proportional to $\dot{M}_{wind}$. Let $\tau_{max}$ denote the value of $\tau$ at the flare maximum. Then we can write: $\tau = \tau_{max} (\dot{M}-\dot{M_0)}/\dot{M}_{max}$, where $\dot{M}_0$ is the accretion rate at $t=0$. According to the meaning of this formula, the value $\tau$ obtained from it represents the increment in the optical thickness of the disk wind caused by an increase in the accretion rate.

Taking this into account, the theoretical light curve has the form: $I = I_{tot} e^{-\tau}+I_{sc}$, where $I_{tot}$ is the intensity of the star’s radiation during the accretion burst, and $I_{ sc}$ is the scattered light. Since we do not know exactly how the accretion flare is transformed into an optical one, for simplicity we will assume: $I_{tot} = I_\star \cdot \dot{M}/\dot{M}_{0}$ and $ I_{sc} =0.01I_{tot}$. When calculating the light curves, it was assumed that $\tau_{max} = 5$.

\begin{figure*}[ht!]
\centering \includegraphics[width=0.8\textwidth]{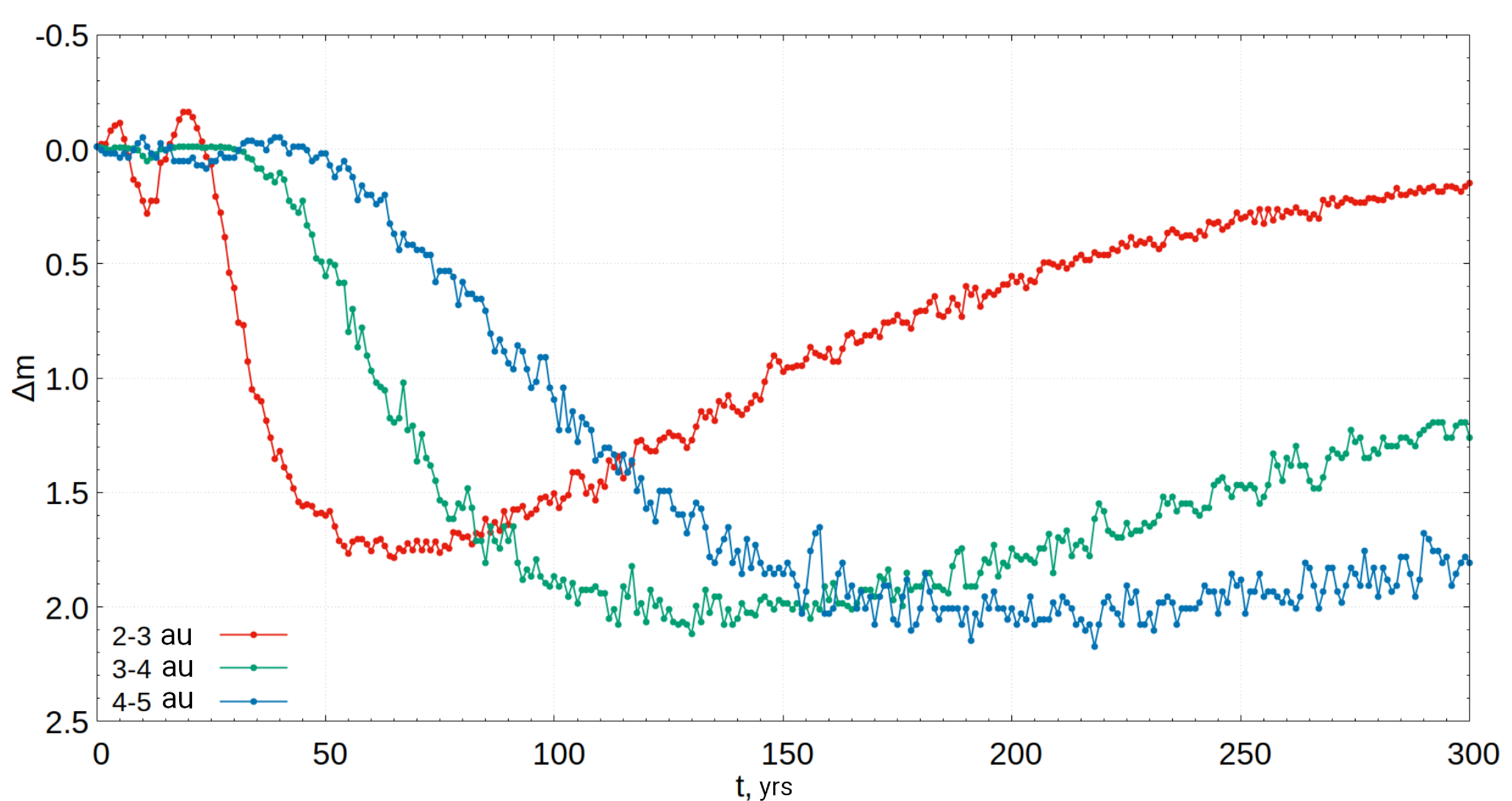}
\caption{\normalsize Theoretical light curves of a star eclipsed by a disk wind for models with parameters $L=0.5$, $I=45^\circ$, $\Delta\phi=6^\circ$ and three ranges $r_0-r_1$ (listed in lower left corner). }\label{fig2}
\end{figure*}

\begin{figure*}[ht!]
\centering \includegraphics[width=0.8\textwidth]{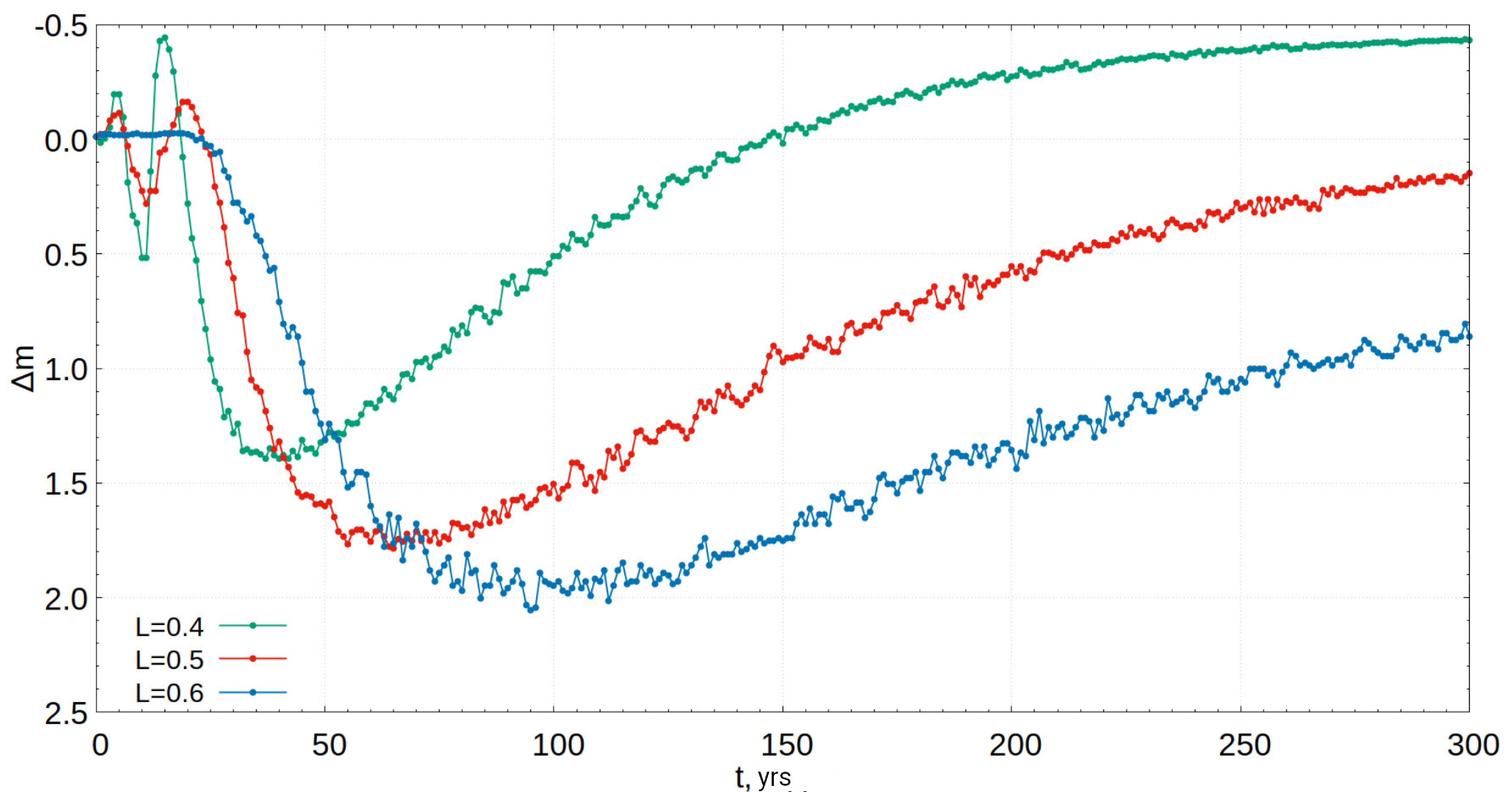}
\caption{\normalsize The same as in Fig.~\ref{fig2} for models with parameters $r_0=2$~au, $r_1=3$~au, $I=45^\circ$, $ \Delta\phi=6^\circ$ and three values of the parameter $L$ (indicated in the lower left corner).}\label{fig3}
\end{figure*}

\begin{figure*}[ht!]
\centering \includegraphics[width=0.8\textwidth]{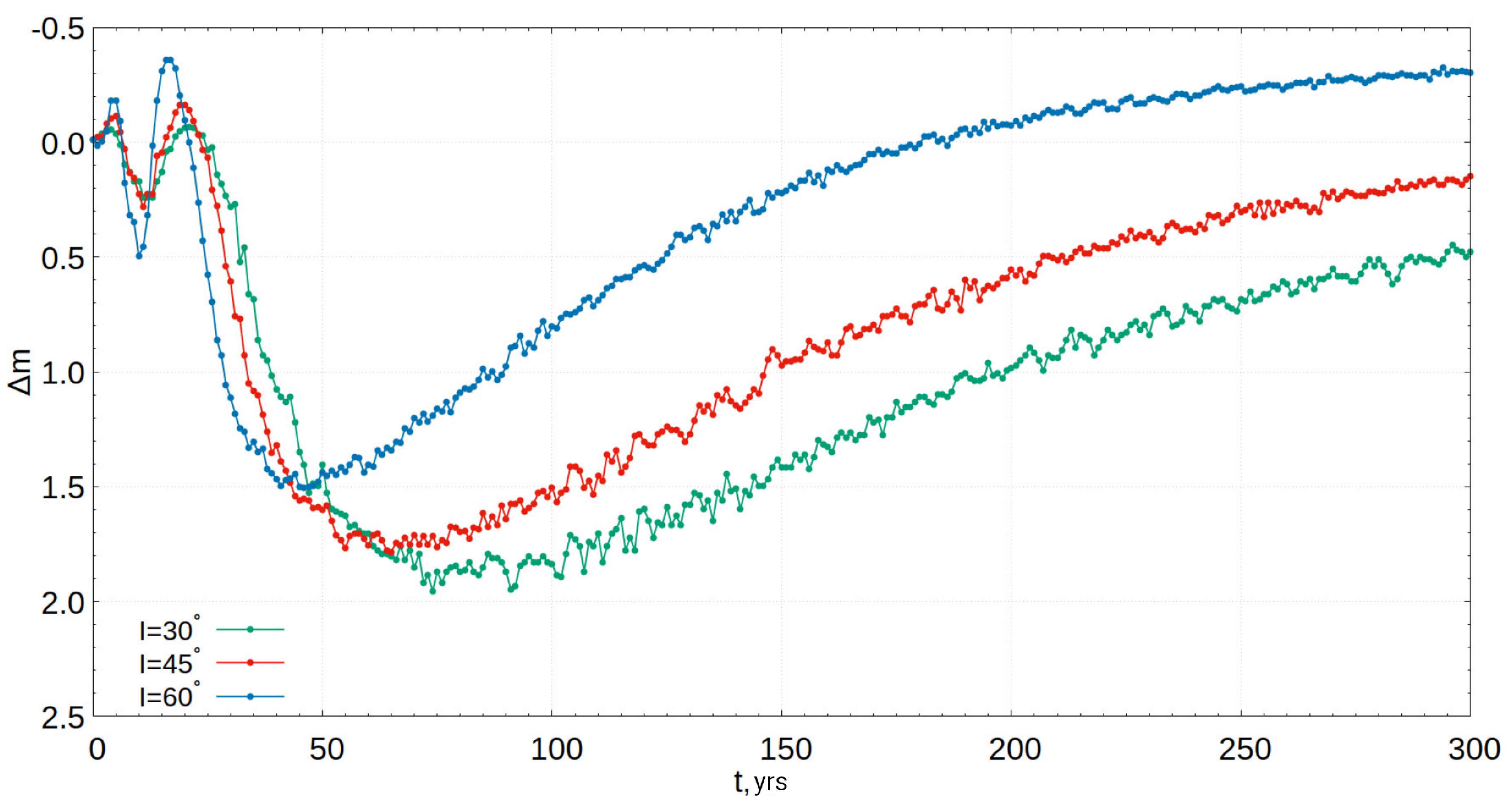}
\caption{\normalsize The same as in Fig.~\ref{fig2} for models with parameters $r_0=2$~au, $r_1=3$~au, $L=0.5$, $\Delta\phi=6^\circ$ and three values of inclination angle $I$ (indicated in the lower left corner).}\label{fig4}
\end{figure*}

The parameters of the calculated models with different characteristics of the clump are given in the Table~\ref{params}. The last two columns estimate the time intervals $\Delta t_d$, during which the brightness decreases, and $\Delta t_{min}$ is the time interval from the beginning of the brightness decrease until the return to the initial value.

\begin{table}[t]
\centering
	\caption{Model parameters} \label{params}
\begin{tabular}{|c|c|c|c|c|c|c|} \hline\hline
$r_0-r_1$ &$L$ &$I$ &$m$ &$\Delta\phi$ &$\Delta t_d$ & $\Delta t_{min}$ \\ 
au & & $^\circ$ & $M_{Jup}$& $^\circ$& yrs & yrs \\ \hline
2-3 &0.5& 45 & 0.3 &6 &35 & $355$ \\ \hline
3-4 &0.5& 45 & 0.3 &6 &60 & $>500$ \\ \hline
4-5 &0.5& 45 & 0.3 &6 &112 & $>500$ \\ \hline
2-3 &0.4& 45 & 0.3 &6 &15 & $130$ \\ \hline
2-3 &0.6& 45 & 0.3 &6 &45 & $>500$ \\ \hline
2-3 &0.5& 30 & 0.3 &6 &55 & $>500$ \\ \hline
2-3 &0.5& 60 & 0.3 &6 &25 & $165$ \\ \hline
2-3 &0.5& 45 & 0.3 &4 &35 & $360$ \\ \hline
2-3 &0.5& 45 & 0.3 &8 &35 & $340$ \\ \hline
2-3 &0.5& 45 & 0.1 &6 &60 & $>500$ \\ \hline
2-3 &0.5& 45 & 0.5 &6 &25 & $127$ \\ \hline
\end{tabular}
\end{table}

Theoretical light curves are shown in the figures~\ref{fig2}--\ref{fig4}. It can be seen that as the distance between the disturbance and the star decreases, the brightness decreases faster, and the minimum phase becomes shorter (Fig.~\ref{fig2}). In a similar way, the light curve is affected by a decrease in the parameter $L$ (Fig.~\ref{fig3}) and an increase in the angle of $I$ (Fig.~\ref{fig4}). Calculations have shown that variations in the $\Delta\phi$ parameter within $4^\circ$--$8^\circ$ do not have a noticeable effect on the behavior of the light curve.

\newpage
\subsection{Light curve of CQ~Tau}\label{sec:res2}

\begin{figure*}[ht!]
\centering \includegraphics[width=0.8\textwidth]{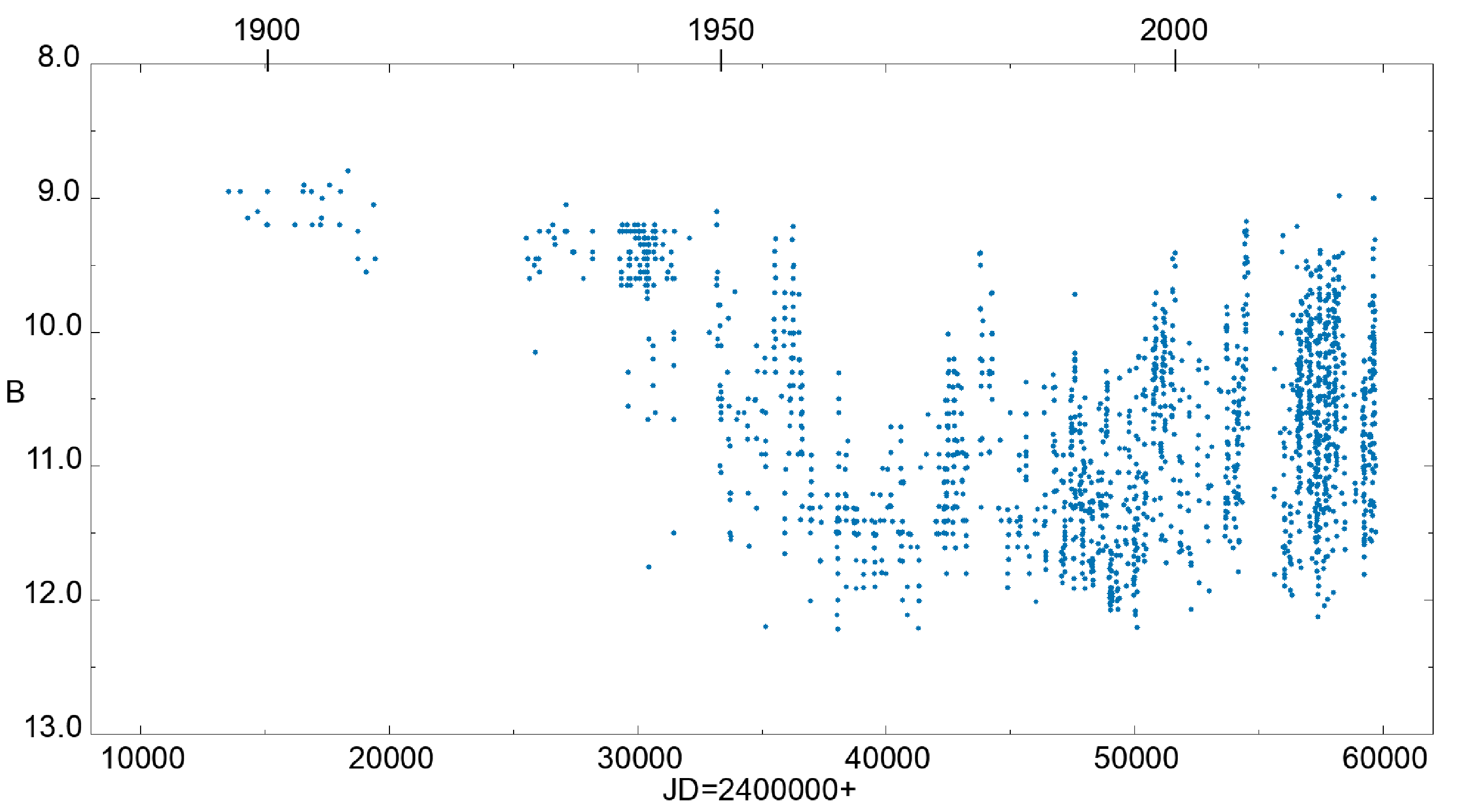}
\caption{\normalsize Historical B-band light curve of CQ~Tau from data~\citet{2023Ap.....66..235G}.}\label{CQTau}
\end{figure*}

Figure~\ref{CQTau} shows the light curve of CQ Tau in the B band over a time interval of $\sim 125$ years. It can be seen that the photometric activity of the star changed dramatically in the middle of the last century: the bright state of the star, accompanied by small flux fluctuations, ended with a sharp drop in brightness caused by a strong increase in circumstellar extinction. This deep minimum has a complex shape and has been going on for more than half a century~\citep{1993Ap.....36...31M,2005Ap.....48..135S,2023Ap.....66..235G}. A recent periodogram analysis of the photometric activity of the star \citep{2023Ap.....66..235G} confirmed the presence of the previously suspected \citep{2005Ap.....48..135S} period of $10$ years~\citep[see also][]{2022MNRAS.515.6109H}. From Fig.~\ref{CQTau} it is clear that the star gradually returns to its original bright state, undergoing frequent and deep dimming of brightness. They indicate that the star is still surrounded by a large amount of highly inhomogeneous circumstellar dust, which, despite the large inclination of the inner disk to the line of sight (about 40$^\circ$ see below), is capable of shielding the star from time to time from the observer.

\begin{figure*}[ht!]
\centering \includegraphics[width=0.8\textwidth]{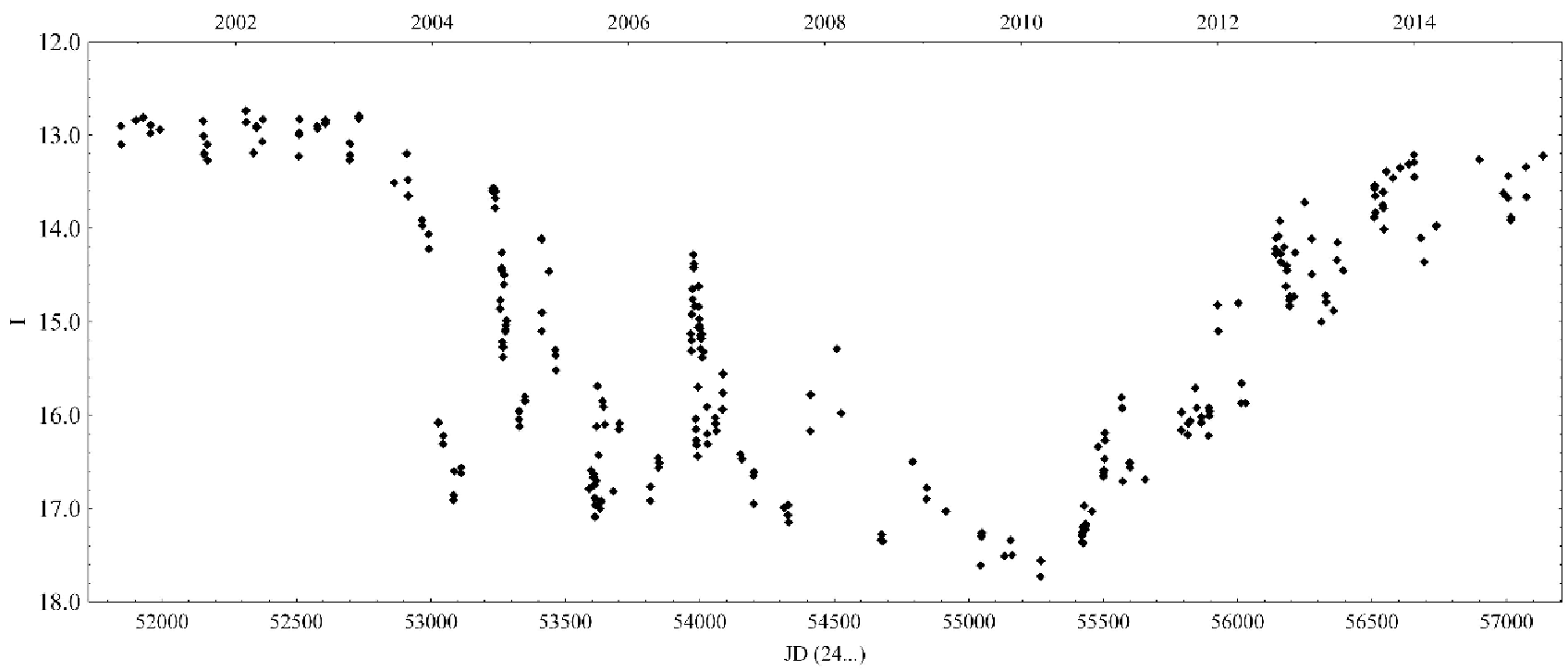}
\caption{\normalsize Light curve of V1184 Tau in band I from ~\cite{2015A&A...582A.113S}; A\&A, A113, 2015, reproduced with permission $\copyright$ ESO.}\label{V1184}
\end{figure*}

Approximately the same shape of a deep minimum lasting about $10$ years was observed~\citep{2015A&A...582A.113S} in the T~Tauri star V1184~Tau, the light curve of which is presented in Fig.~\ref{V1184}. The general similarity between the theoretical and observed light curves presented in figures~\ref{fig2}--\ref{fig4} and \ref{CQTau},\ref{V1184} suggests that in both cases a strong and prolonged decrease in the brightness of these stars was initiated by a disturbance in their circumstellar disks, the cause of which could be the fall of a massive clump onto a disk in the vicinity of the star. It should be noted that currently CQ Tau is characterized by a rather high accretion rate: $\dot M = 1.12 \cdot 10^{-7} M_\odot$ per year~\citep{2011AJ....141...46D} . At this accretion rate, the dusty disk wind can be opaque to the optical radiation of the star, even when the disk is tilted to the line of sight at about $40^\circ$~\citep{2024AstL...T}. Disturbances caused by the orbital motion of the companion\footnote{There are reasons to assume that its orbit has a large eccentricity \citep{2023Ap.....66..235G}}, lead to periodic variations in the amplitude of the eclipses, which are clearly visible on the light curve of the star (Fig.~\ref{CQTau}). 

A similar sharp drop in brightness occurred in the star AA Tau in 2011, and the star is still in this state~\citep{2013A&A...557A..77B, 2021AJ....161...61C}. Another T~Tauri star, RW~Aur, has had photometric activity that has changed greatly in recent years: the star began to experience deep and prolonged minima~\citep{2015IBVS.6143....1S,2015A&A...577A..73P,2016A&A...596A..38F,2019MNRAS.482.5524D}. Such large-scale dimming of the brightness of young stars indicates the appearance in their immediate vicinity of an additional portion of matter, which stimulated the rise of dust above the disk, including due to an increase in the disk wind.
\section{Discussion}
The dusty disk wind can be not only a source of circumstellar extinction, which causes a weakening of the optical brightness of the star at small angles of the disk inclination to the line of sight, but also a source of additional IR radiation~\citep{2012ApJ...758..100B}. As a result, an interesting phenomenon can be observed when the optical brightness of a star decreases, and the IR radiation increases. This phenomenon has indeed been observed in some UX~Ori type stars. In particular, an increase in brightness in the K band during the optical minimum was observed for the star V1184~Tau~\citep{2009AstL...35..114G} and was interpreted as an increase in the disk wind. A similar picture was observed~\citet{2015IBVS.6143....1S} for the star RW~Aur and was interpreted in the same way.  

In addition to the ten-year photometric period in changes in the star's brightness, the existence of a companion in the vicinity of CQ~Tau is also evidenced by the results of interferometric observations of the star in the millimeter range ~\citep{2019MNRAS.486.4638U,2021A&A...648A..19W}, which showed the presence of an extended cavity in the central parts of the circumstellar disk\footnote{Such cavities are known to form in disks as a result of tidal disturbances caused by the orbital motion of companions~\citep{1994ApJ...421..651A}}. According to interferometric observations in the near-IR region of the spectrum, the inner part of the CQ~Tau disk is inclined at an angle $i = 48\pm 5^\circ$~\citep{2004ApJ...613.1049E} relative to the plane of the sky, while according to observations in the millimeter range, the outer disk is inclined at an angle of $35^\circ$~\citep{2008A&A...488..565C,2019MNRAS.486.4638U}. As is known, a similar picture is observed in the central region of the circumstellar disk $\beta$~Pic~\citep{1995AAS...187.3205B} and is explained by the existence of a planet whose orbit is inclined relative to the plane of the outer disk~\citep{2009A&A...506..927L,2012A&A...542A..41C}. The different inclinations of the inner and outer disk indicate strong disturbances in the process of its formation and evolution.
\section{Conclusion}\label{sec:concl}
Thus, the fall of a massive clump onto the disk can cause a luminosity burst of the FU~Ori type when observing the disk from the polar directions \citep{2023ApJ...953...38D}, whereas with a slight inclination of the disk to the line of sight such an event can cause a deep and long lasting reduction of brightness. This dualism of the FU~Ori phenomenon emphasizes the importance of the orientation of the circumstellar disk relative to the direction towards the observer when describing the observed phenomena. In the case of UX~Ori type stars, to which CQ~Tau belongs, the conclusion about the small angle of inclination of the disk to the line of sight was made based on observations of high linear polarization of stars in deep brightness minima~\citep{1991Ap&SS.186..283G}. For CQ~Tau itself, such observations were obtained~\citet{1990AZh....67..812B}. It is interesting that the assumption that V1184~Tau belongs to the family of UX~Ori type stars was made in~\citet{1997AJ....113.1395A} long before the star began its deep minimum brightness in 2004, shown in Fig.~\ref{V1184}. The basis for this assumption was spectral observations, which showed that the emission in the $H_\alpha$ line in the spectrum of the star has a two-component profile, expanded by the rapid rotation of the emitting gas. Previously, based on statistical analysis, it was shown~\citep{1996ARep...40..171G} that such $H_\alpha$ line profiles are characteristic specifically of UX~Ori type stars, which was recently confirmed in~\citet{2018A&A...620A.128V} based on more extensive statistics.

It should be noted that in the case of T~Tauri stars with intense accretion, determining the angle of inclination of the disk to the line of sight based on interferometric observations in the near-IR region of the spectrum may contain a systematic error. It caused by the fact that, unlike the disk, which is, to a good approximation, a two-dimensional structure, the disk wind, which is a potential source of IR radiation, is a three-dimensional object, and when projected onto the plane of the sky, its image may differ noticeably from the image of the disk.
\newpage
{\bf Acknowledgments}
The authors thank an anonymous reviewer for helpful comments.
Simulations were performed using the resources of the Joint SuperComputer Center of the Russian Academy of Sciences — Branch of Federal State Institution ``Scientific Research Institute for System Analysis of the Russian Academy of Sciences''\footnote{\href{https://www.jscc.ru/}{https://www.jscc.ru/}}~\cite{2019LG..40..1835}. 
\newpage
 \bibliography{biblio}{}
\bibliographystyle{aasjournal} 
\end{document}